\documentclass[preprint,1p,times,authoryear]{elsarticle}

\usepackage{graphicx} 
\graphicspath{{Pictures/}}

\usepackage{caption}
\usepackage{multirow}
\usepackage{subcaption} 
\usepackage{booktabs} 
\usepackage{amsmath} 
\usepackage{amssymb}
\usepackage{color}
\usepackage{enumitem}
\journal{International Journal of Plasticity}

\begin{document}
\begin{frontmatter}

\title{Applied Machine Learning to Predict Stress Hotspots II: Hexagonal close packed materials}%

\author[cmu]{Ankita Mangal}
\ead{mangalanks@gmail.com}
\author[cmu]{Elizabeth A. Holm\corref{cor1}}
\ead{eaholm@andrew.cmu.edu}

\cortext[cor1]{Corresponding author}
\address[cmu]{Department of Materials Science and Engineering, Carnegie Mellon University, 5000 Forbes Ave, Pittsburgh, PA 15213, USA}

\begin{abstract}
Stress hotspots are regions of stress concentrations that form under deformation in polycrystalline materials. We use a machine learning approach to study the effect of preferred slip systems and microstructural features that reflect local crystallography, geometry, and connectivity  on stress hotspot formation in hexagonal close packed materials under uniaxial tensile stress. We consider two cases: one without any preferred slip systems with a critically resolved shear stress (CRSS) ratio of 1:1:1, and a second with CRSS ratio 0.1:1:3 for basal: prismatic: pyramidal slip systems. Random forest based machine learning models predict hotspot formation with an AUC (area under curve) score of 0.82 for the Equal CRSS and 0.81 for the Unequal CRSS cases. The results show how data driven techniques can be utilized to predict hotspots as well as pinpoint the microstructural features causing stress hotspot formation in polycrystalline microstructures.

\end{abstract}


\end{frontmatter}

\section{Introduction}
In polycrystalline materials, an applied stress is distributed inhomogeneously, resulting in stress concentrations, termed stress hot spots. An important mechanism for ductile fracture in metals and their alloys is by the growth and coalescence of microscopic voids, which nucleate near stress hotspots (\cite{Rimmer1959}). In face centered cubic (fcc) materials under uniaxial tensile deformation, stress hotspots tend to form near microstructural features and usually occur in textures corresponding to maxima in the Taylor factor (\cite{Rollett2010a, Mangal2017b}). Crystalline anisotropy, which determines the "hard" and "soft" directions; also plays an important role. In FCC materials where these directions change between elastic and plastic regimes, the elastic hotspots can become plastic coldspots (\cite{Lebensohn2012}). 

\begin{figure}[!t]
\centering
\includegraphics[width = 0.4\textwidth]{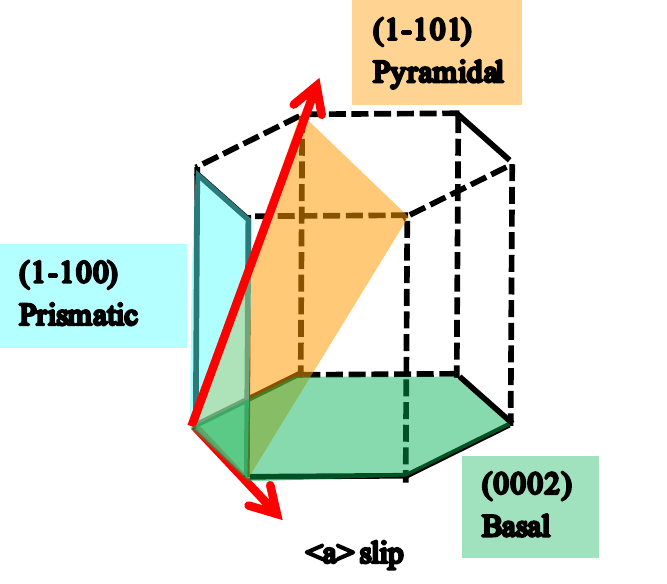}
    \caption{Schematic of the different slip systems in a hexagonal close packed structure:  basal $\{0001\}[1120]$, prismatic $\{1010\}[112]$ and pyramidal $< c + a >$. 
    When the tensile axis lies in the ($10\bar{1}1$) pyramidal plane, the Schmid factor of prismatic $<a>$ slip is higher than the basal $<a>$ slip.}
    \label{fig:EqualCRSSslips}
\end{figure}

The elastic/ plastic behavior of hexagonal close packed (HCP) materials is more complex due to the inherent anisotropy of a non-cubic crystal structure. As shown in Figure \ref{fig:EqualCRSSslips}, HCP materials deform plastically by slip on 4 slip systems: basal $\{0001\}[1120]$, prismatic $\{1010\}[112]$, pyramidal $< a >$ $\{1101\}[1120]$ and pyramidal $< c + a >$, each with different critical resolved shear stress (CRSS) values (\cite{Thornburg1975}). (Deformation twinning also adds to the complexity but has been ignored in this work.) Deformation textures developed in HCP materials vary due to the unique slip and twinning systems that are activated based on the c/a ratio and the critically resolved shear stress (CRSS) of basal and non basal slip modes.

To understand polycrystal plasticity and texture development in terms of single crystals, the concept of the single crystal yield surface (SCYS) was developed. The SCYS determines the shears that are activated in a grain and depends on the CRSS ratios between deformation modes, as well as the stress state. The SCYS has been analyzed and derived in detail for BCC materials in \cite{Orlans-Joliet1988}, for FCC materials in \cite{Rocks1983} and HCP materials in \cite{Tome1985}. \cite{Chin1970} showed that the SCYS is topologically invariant in certain domains of CRSS ratios, and leads to a simplified analysis of deformation when slip modes harden at different rates. 

The CRSS ratio is defined with respect to the basal slip resolved shear strength ($\tau_{basal}$) as:

\begin{equation}\label{eq:CRSS_ratio}
\begin{aligned}
CRSS Ratio = 
= \frac{\tau_{prismatic}}{\tau_{basal}}:1:\frac{\tau_{pyramidal}}{\tau_{basal}}
\end{aligned}
\end{equation}
where $\tau_{prismatic}$ and $\tau_{pyramidal}$ are the CRSS of prismatic and pyramidal slip systems respectively. Even if the CRSS of a mode is very high, it might be activated to complete the yield surface to achieve the 5 independent slip modes required by the Taylor and Von-Mises criteria, resulting in a highly anisotropic macroscopic response (\cite{taylor1938plastic, Piehler2009}). The situation is worsened by the need to satisfy compatibility and equilibrium conditions between neighboring grains, and results in the material selecting a spatially inhomogeneous solution to accommodate the macroscopic boundary conditions.

Changing the texture of the material will have the same effect of making some slip systems more favorable than others. Hence in order to understand the evolution of stress hotspots, it is necessary to look into a combination of all these variables: texture, grain shape, c/a ratio, CRSS ratios, slip hardening, twinning, temperature and stress state. In this work, we keep the temperature constant, and uniaxial tensile deformation is constrained to occur only by 3 slip modes: prismatic, basal and pyramidal $<c+a>$ without any twinning or anisotropic slip hardening. The microstructure consists of equiaxed grains and the c/a ratio is fixed. Thus, we can vary the CRSS ratio and crystallographic texture to analyze their impact on stress hotspot formation.

Machine learning (ML) techniques are gaining popularity and have been applied successfully to various fields  (\cite{LeCun2015,bose2001business, lavecchia2015machine, McMahan2013a, Mangal2017a}) to gain insights and relationships between features or attributes of different kinds. These techniques are finding their way into the materials science domain (\cite{Rajan2015, Fedorov2017a, Gomez-Bombarelli2016}), in areas such as molecular informatics (\cite{Yao2017}), predicting deformation twinning based on the local structure (\cite{Orme2016}) and predicting phase diagrams (\cite{meredig2014combinatorial}). In a companion paper, we have used ML methods to analyze stress hotspots in FCC materials (\cite{Mangal2017b}). Our model was based on local microstructural features that describe the crystallography ( Euler angles, Schmid factor, misorientations) and geometry (grain shape, grain boundary types). The target to was predict whether a grain becomes a stress hotspot based on a feature vector $\mathbf{X}$ whose components are the local microstructural descriptors. In this work, we extend this approach to study stress hotspots in HCP materials as a function of texture and compare them among two different HCP materials: an ideal Equal CRSS ratio case where the CRSS ratio is $1:1:1$ and an Unequal CRSS ratio case of $0.7:1:3$. We then compare the performance of machine learning models and delineate the microstructural features that contribute the most in predicting stress hotspots. 

\section{Methods}
\subsection{Dataset Generation}
\begin{figure*}[t]
\centering
            \includegraphics[width=\textwidth]{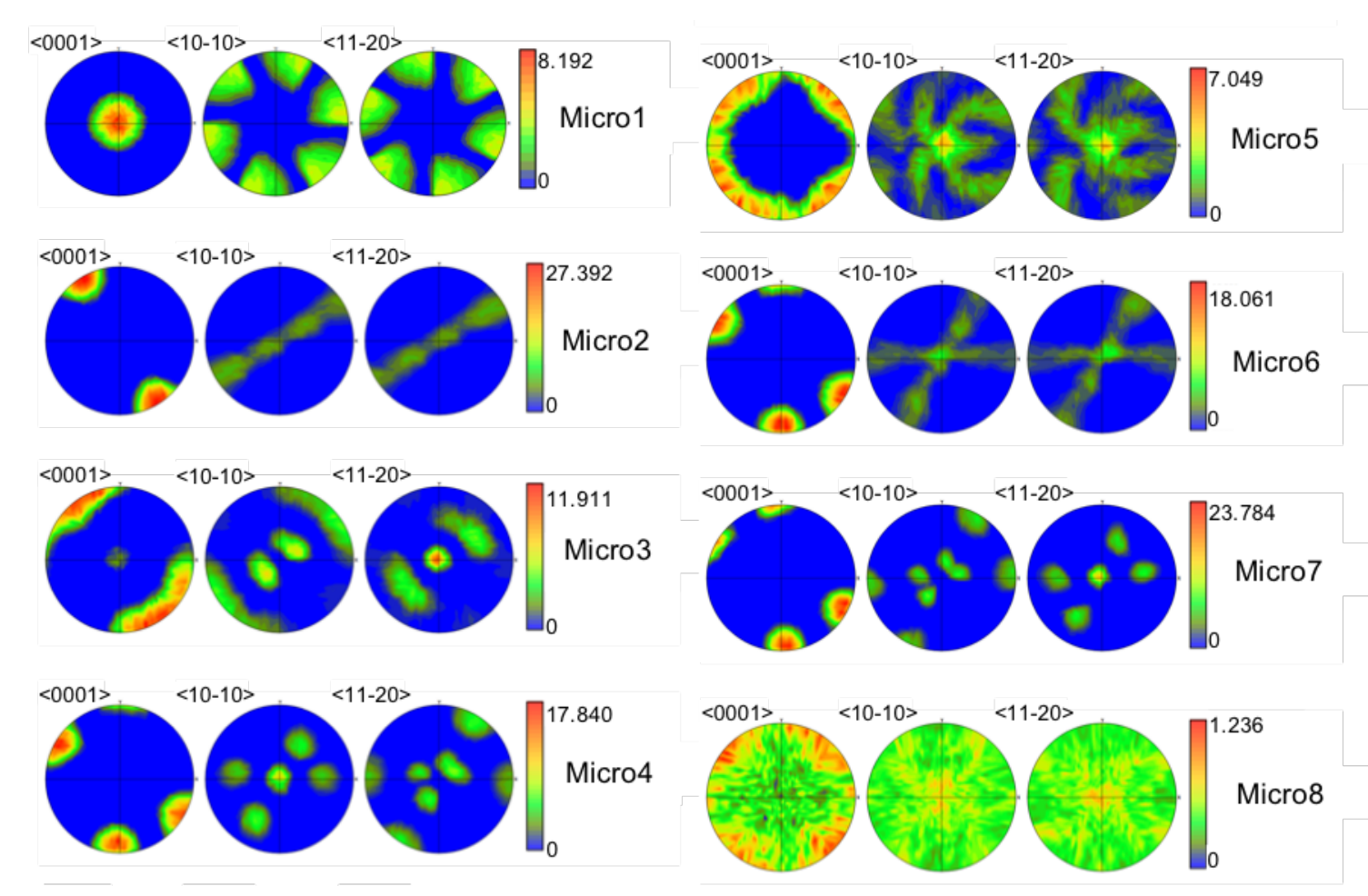}
            \caption{Representative textures for 8 different HCP textures, the corresponding scale bars show the texture intensity in MRD }
            \label{fig:HCPtex}
\end{figure*}

We use the Dream.3D package (\cite{Groeber2014}) to generate a dataset of synthetic polycrystalline microstructures with a mean grain size of 2.7 microns consisting of $\sim 5000$ grains each. We study 8 representative textures shown in Figure \ref{fig:HCPtex}. For each representative texture, 9 stochastic microstructure instantiations were created, resulting in $\sim 45000$ grains per texture. The texture intensity for each microstructure instantiation varied from weak ($<$5 MRD)  to strong  ($>$30 MRD), where MRD (multiples of random density) denotes the intensity of a crystallite orientation with respect to a randomly textured material.

The microstructures were then discretized on a $128\times128\times128$ grid to facilitate the use of EVPFFT (elasto-viscoplastic fast Fourier transform): an image based crystal plasticity formulation from\cite{Lebensohn2012} to simulate uniaxial tensile deformation. The constitutive model parameters for HCP materials represent a general alpha-titanium alloy and are summarized in the supplementary material (table \ref{HCPelastic1}). The EVPFFT model uses the \textbf{Voce hardening} law (\cite{voce1955practical}) to model strain hardening as follows:
\begin{equation}
\tau^{s}(\Gamma) = \tau^{s}_{0} +(\tau^{s}_{1}+\theta^{s}_{1}\Gamma)\left(1-exp\Big(
-\Gamma\mid \frac{\theta^{s}_{o}}{\tau^{s}_{1}}\mid\Big)\right)
\end{equation}

where for a given slip system s, $\tau_0$ is the initial yield stress and and $\theta_0$ is the initial hardening rate. $(\tau_0 + \tau_1)$ is the back-extrapolated stress and $\theta_1$ is the asymptotic hardening rate. $\Gamma$ is the accumulated shear in the grains. The Voce hardening parameters were extracted by fitting the VPSC code generated stress-strain curves (\cite{Lebensohn1993}) to the experimentally obtained curve as shown in \ref{AppendixA}. 

\subsection{Microstructural Descriptors}
The dataset consists of voxel-wise representation of the stresses in each microstructure. The spatially resolved stress field is then averaged grain-wise to minimize the impact of numerical artifacts and small-scale fluctuations. The resultant Von Mises stress distribution is then thresholded above the $90^{th}$ percentile to designate stress hotspots following the same procedure as in \cite{Mangal2017b}. This results in $10\%$ of the grains designated as stress hotspots.

Stress distribution in a microstructure is affected by crystallography as well as grain neighborhood and geometry. Hence we develop microstructural features describing the crystallography, geometry and connectivity of grains; and use these as input features to a machine learning algorithm that predicts whether a grain is hot or not. We have developed a number of microstructural features in \cite{Mangal2017b}. Along with those features, we include additional HCP material specific features describing the crystallography and geometry to be used in this paper. \ref{Featurenames} lists the acronyms and descriptions of the features used in this work. 

The crystallographic descriptors include distance from inverse pole figure corner, which quantifies a grain's orientation with respect to the $[001], [010]$ and $[100]$ directions in the sample frame; features quantifying the misorientation between a grain and it's neighbors, and Schmid Factors for each of the basal, prismatic and pyramidal slip systems for each grain. Due to the inherent anisotropy in HCP materials, the orientation of the HCP c-axis with respect to the tensile axis is also a good descriptor. 

The geometry based descriptors include shape averaged Euclidean distance from special points such as grain boundaries, triple junctions and quadruple points.  Features based on grain shape include grain size, equivalent diameter, volume, number of contiguous neighbors, number of neighbors, grain aspect ratio and surface area to volume ratio. 

We now have datasets for \textit{Equal CRSS} and \textit{Unequal CRSS} materials consisting of grain-wise labels denoting stress hotspots, and grain-wise features for each microstructure. Each dataset has 72 microstructures (8 representative texture kinds and 9 microstructures per texture). Although both the Equal and Unequal CRSS ratio data sets represent HCP materials with the same c/a ratio, their constitutive parameters are different, so they fundamentally represent different materials. Hence a machine learning model is built for each case to predict whether a stress hotspot forms in a given grain. 

\subsection{Machine Learning Methods}
Since the stress distribution in a microstructure is impacted by a complex interplay of crystallography, geometry and connectivity, we want to build a predictive model which minimizes the assumptions about which features cause hotspot formation. Machine learning models present this opportunity by providing a statistical framework to create connections between the target to be predicted (hotspot) and the features describing it (\cite{Mitchell1997}). This is achieved by training a model over training data and evaluating the model performance on validation data which gives us an estimate of the model performance on unseen data; also known as the generalization error. When the amount of data available is small, k-fold cross validation is used to get the model performance estimate. In this technique, the available dataset is divided into k subsamples; the model is trained on (k-1) subsamples and validated on the $k^{th}$ subsample. This process is repeated k times to get the validation error in each fold, which is then averaged to get the model generalization error estimate. 

The generalization error consists of "bias" and "variance" corresponding to the need for a more complex model and the need for more data respectively. These insights into the generalization error can be obtained by the use of \textit{learning curves}. A learning curve is the plot of training and validation performance of the model as more and more data is used to train the model. A gap between the training and validation performance in the learning curve signals the need for additional data to minimize the generalization error due to variance. On the other hand, if the training and generalization errors converge, but are both low, it means that bias predominates and we need to introduce a more complex algorithm and more descriptive features.

In this paper, we utilize a decision tree based model known as the random forest (RF) algorithm (\cite{Breiman1996, Breiman2001}) to build our classification model.  RF models are very fast and easy to fit, can handle all kinds of features (numerical, categorical) and deal with missing features or data effectively. RF models also provide the importance of the predictor variables (\cite{Breiman1996}), although correlations between features can cause inaccuracies in feature importance rankings (\cite{Gregorutti2016}). The model hyper-parameters include the number of decision trees, number of features and the depth of the decision trees. The hyper parameters are chosen using a random grid search by comparing the cross validation performance. The model was implemented using the Scikit-learn library in Python (\cite{pedregosa2011scikit}). More details of the RF model are included in the companion paper \cite{Mangal2017b}.

The aim is to classify the grains as hot or normal, which is a binary classification problem. Using \textit{accuracy} as the model performance metric is not suitable in this case due to the imbalance between the two classes; only $10\%$ of the grains are designated as hot. Hence we use the area under the receiving operator characteristic curve (AUC) metric to compare model performance (\cite{auc}). If the classifier is no better than random guessing, the AUC will be around 0.50. A good classifier has an $AUC \sim1$.

Since varying the texture also has an impact on the location of stress hotspots (\cite{Mangal2017b}), we build two kinds of models for each case: 
\begin{itemize}[leftmargin=*]
    \item \textbf{Partition models}: For each case and texture class; a different random forest model is trained. Model performance is reported using k-fold cross validation calculated as the average of the validation performance metrics on each microstructure in a texture class.
    \item \textbf{Mixed-model}: For each case, a single random forest model is trained on all the 72 microstructures. Model performance is reported using k-fold cross validation calculated as the average of the validation performance metrics on two randomly chosen microstructures from each texture class.
\end{itemize}

Finally, we use the FeaLect method (\cite{Zare2013}) to extract feature importances from the dataset, which is then used to derive data driven insights. FeaLect is a state of the art feature selection algorithm that is robust to correlation between the features, and selects the subset of features most highly correlated to the target but least correlated to one another. First the dataset was oversampled to balance the population of the two classes. It was then bootstrapped into 100 subsets. In each random subset,  linear models are fitted using least angle regression (LARS) method  (\cite{Efron2004}) with the regularization strength such that only 10 features are selected in each model. Features are scored on their tendency to be selected in each model. Finally, these scores are averaged to give the feature importance on an absolute scale. We used the R implementation of FeaLect to compute our results (\cite{Zare2015}).

\section{Results and Discussion }
\begin{figure}
	\centering
	\begin{subfigure}[t]{0.49\textwidth}
         \includegraphics[width=\textwidth]{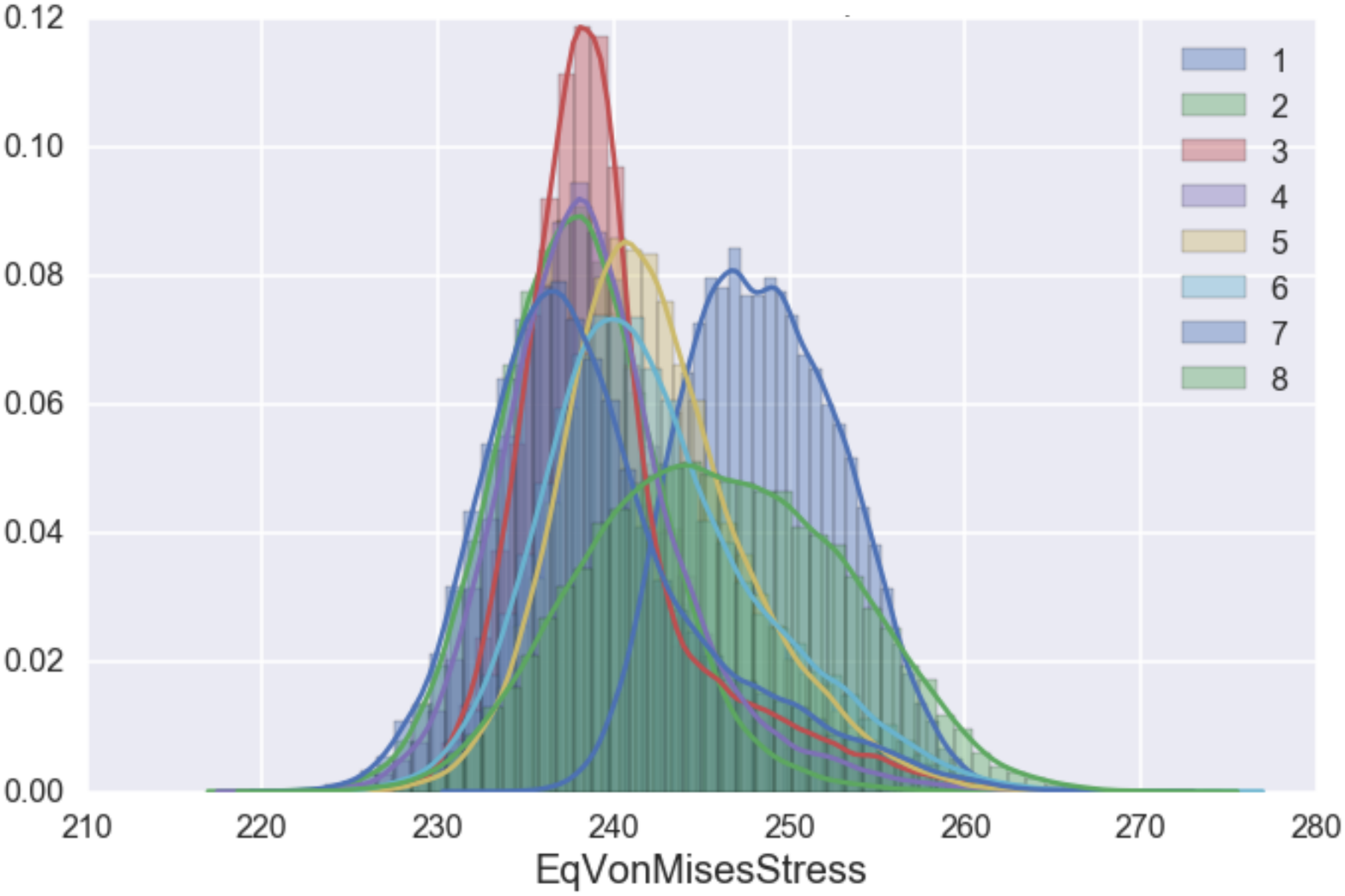}
    	\caption{}
	\label{fig:HcpDatasetStress}
    	\end{subfigure}
    \hspace{\fill}
    \begin{subfigure}[t]{0.49\textwidth}
    \includegraphics[width=\textwidth]{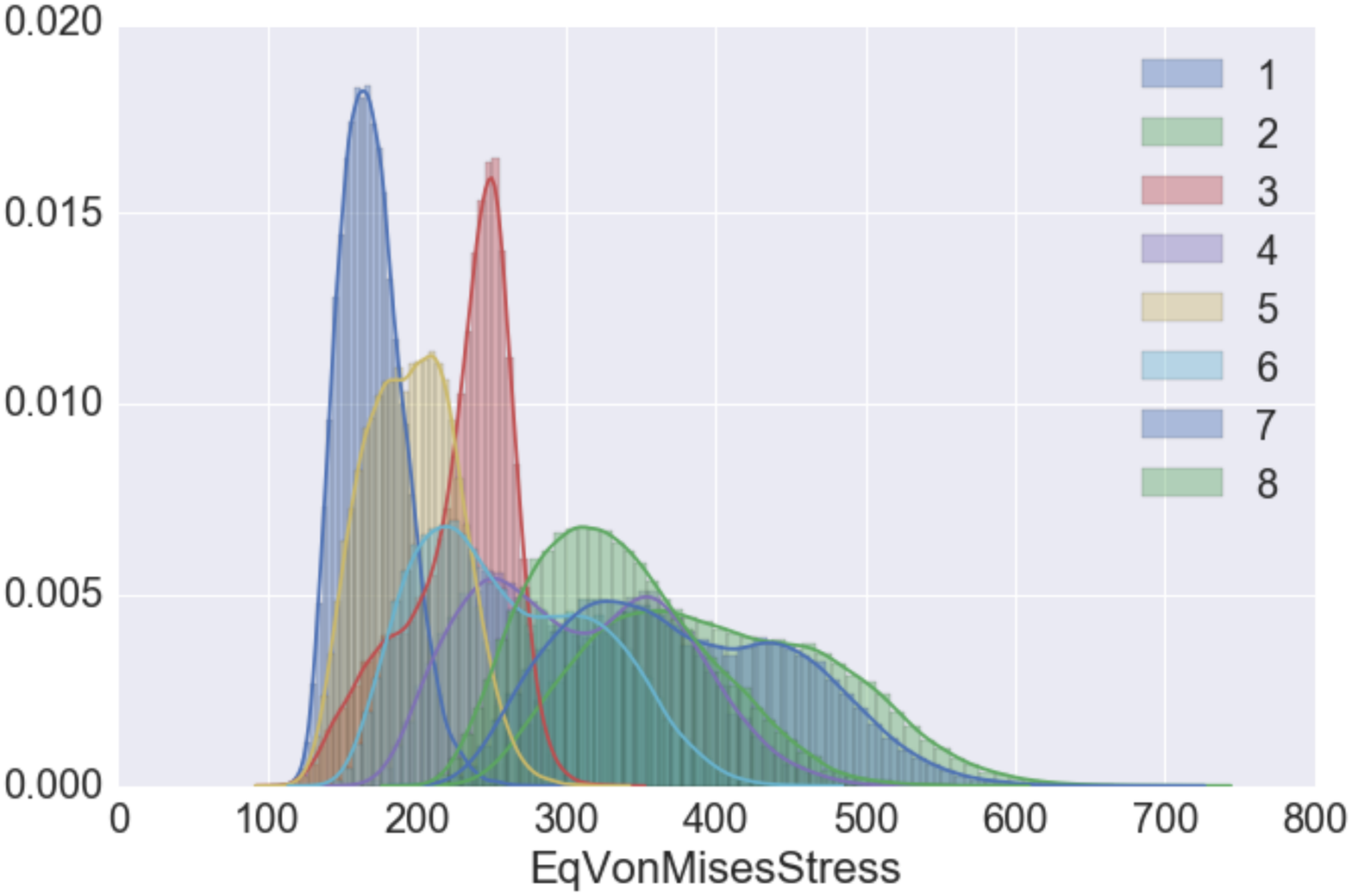}
    \caption{}
    \label{fig:Slipr_stressDist}
    \end{subfigure}
    \\
    \begin{subfigure}[t]{0.49\textwidth}
         \includegraphics[width=\textwidth]{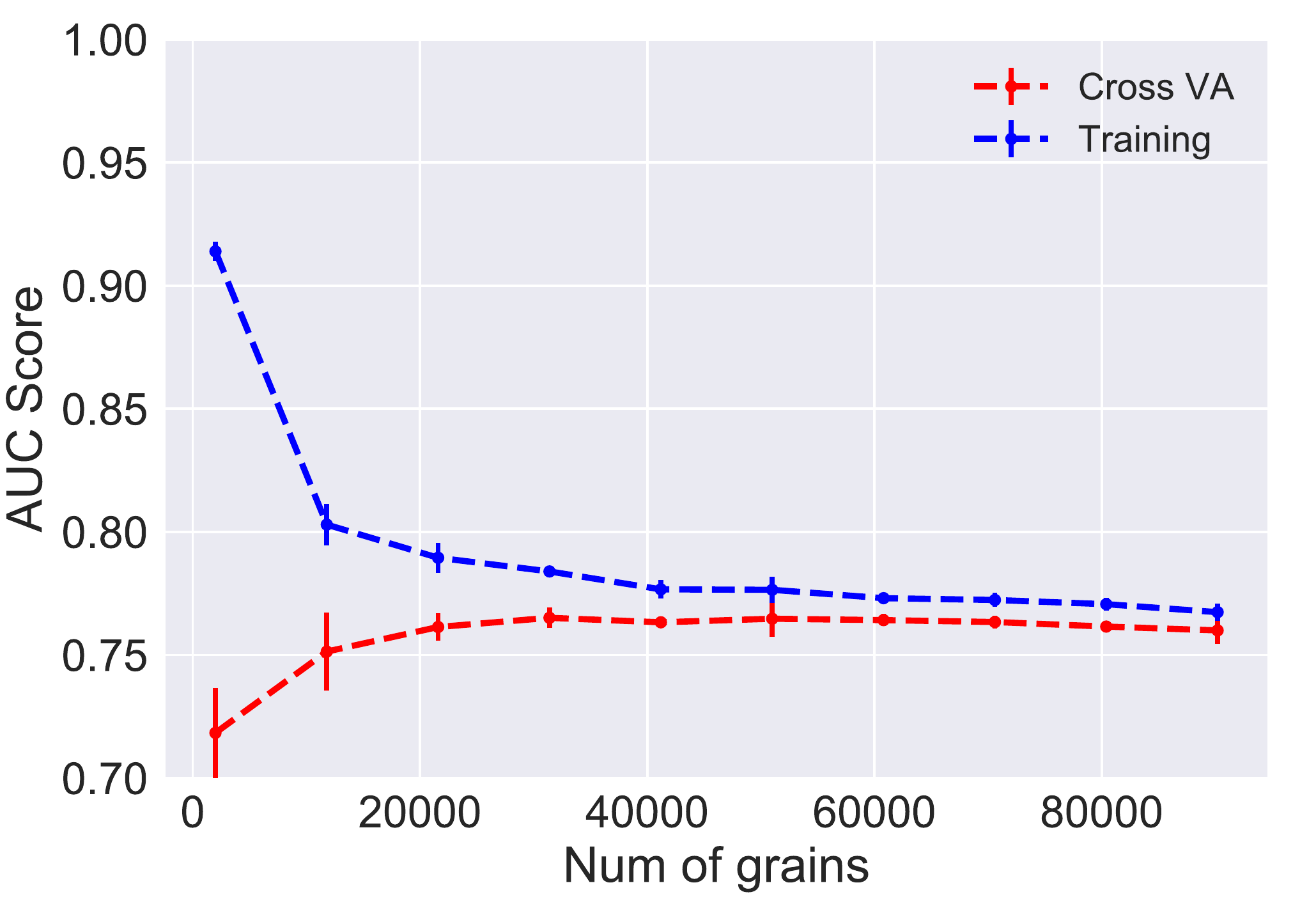}
        \caption{}
        \label{HCP_learning curve}
     \end{subfigure}
    \hspace{\fill}
    \begin{subfigure}[t]{0.49\textwidth}
    \includegraphics[width=\textwidth]{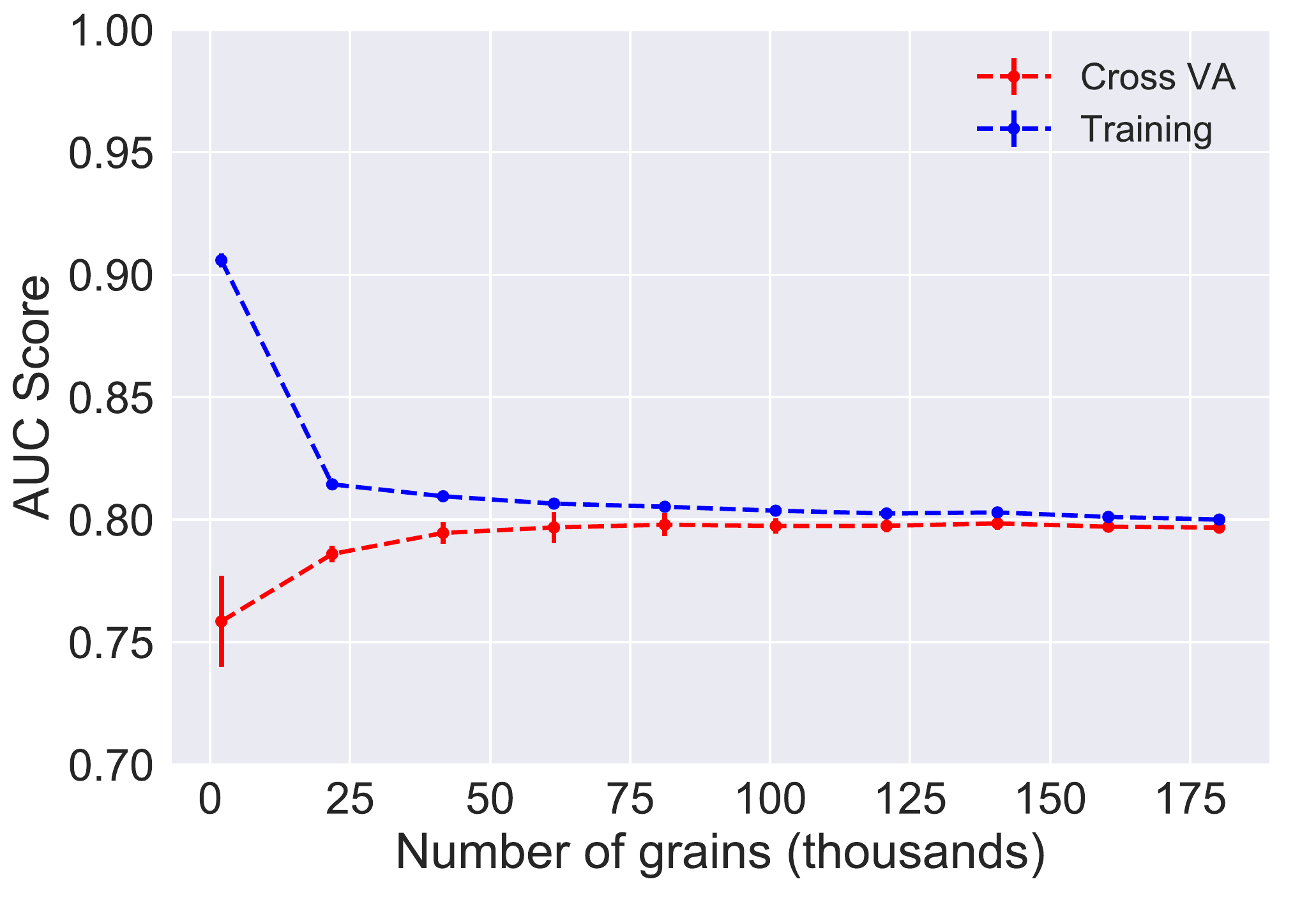}
        \caption{}
        \label{hcpunequal_learning curve}
    \end{subfigure}
    \caption{Histograms of grain averaged stress with different textures in HCP materials with (a) Equal CRSS ratio and (b) Unequal CRSS ratio. The corresponding learning curves for Mixed-Micro model in HCP materials with (c) Equal CRSS ratio and (d) Unequal CRSS ratio}
\end{figure}

For the equal CRSS material, the ratio of basal$<a>$ : prismatic$<a>$: pyramidal$<c+a>$ CRSS is 1:1:1. It is worth noting that this CRSS ratio is not observed in $\alpha$-Ti, and represents an ideal HCP material with isotropic slip systems. Figure \ref{fig:HcpDatasetStress} shows the representative grain averaged stress distribution in each texture class for the Equal CRSS ratio case: the stress distributions are all right tailed. 

For the Unequal CRSS ratio case, uniaxial tensile deformation is simulated with the same microstructure set as the Equal CRSS ratio case, but using different constitutive parameters. The CRSS ratio chosen is basal$<a>$ : prismatic$<a>$: pyramidal$<c+a>$ = 1: 0.7 : 3. This CRSS ratio is selected to better represent $\alpha$-Ti (\cite{Semiatin2001}). It was observed that due to the inhomogeneity in CRSS values, texture heavily influences the macroscopic response. Figure \ref{fig:Slipr_stressDist} shows the grain averaged stress distribution in each texture class. The stress distributions change character between different textures.  

Partition and Mixed random forest models were computed for both the CRSS ratio cases separately. The optimized model hyper-parameters are: max depth=8 and num estimators=1200. Table \ref{AUC_hcp} reports the AUC score using 9 fold cross validation for Partition models and 8 fold cross validation for Mixed models. It is found that Mixed models perform comparably or better than the Partition models for both the datasets. This is a surprising positive result, as it eliminates the need for training different models for each material texture class. The benchmark predictive power (AUC) of the mixed microstructure model is $82.5\pm8.22\%$ and $81.18\pm5.94\%$ in the Equal and Unequal CRSS ratio cases respectively.

The learning curves for the mixed models for the two datasets are shown in figure \ref{HCP_learning curve} and \ref{hcpunequal_learning curve}. The training and validation model performance seem to converge in both cases which means the model performance can be improved by either increasing the feature space, or by using a more complex model algorithm.

\begin{table}
\begin{center}
\caption{Cross validation AUCs (\%) for mixed and partition models in equal and Unequal CRSS ratio case of HCP materials}
\label{AUC_hcp}
\begin{tabular}{*{5}{c}}
\toprule  \multirow{2}{*}{\textbf{Texture kind}} & \multicolumn{2}{c}{\textbf{Equal CRSS}} &  \multicolumn{2}{c}{\textbf{Unequal CRSS}} \\ 
& Partition model AUC & Mixed model AUC & Partition model AUC & Mixed model AUC \\
\midrule
1&$87.47\pm0.67$&87.84&$71.87\pm0.65$&71.73\\
2&$66.20\pm8.65$&77.93&$86.61\pm0.75$&85.78\\
3&$74.89\pm6.44$&90.51&$72.52\pm3.35$&75.94\\
4&$69.89\pm11.82$&78.27&$83.20\pm5.07$&82.89\\
5&$83.22\pm11.98$&89.79&$76.78\pm3.74$&73.72\\
6&$79.89\pm10.03$&86.12&$77.62\pm6.46$&87.61\\
7&$73.39\pm8.52$&64.19&$85.31\pm1.78$&85.43\\
8&$85.48\pm0.51$&85.37&$87.76\pm0.61$&86.22\\
\midrule
All&$77.55\pm7.66$ & $82.50\pm8.22$ &$80.21\pm6.33$ &$81.18\pm5.94$\\
\bottomrule
\end{tabular}
\end{center}
\end{table}

\begin{figure}[!bp]
    \centering
    \includegraphics[width=\linewidth, keepaspectratio]{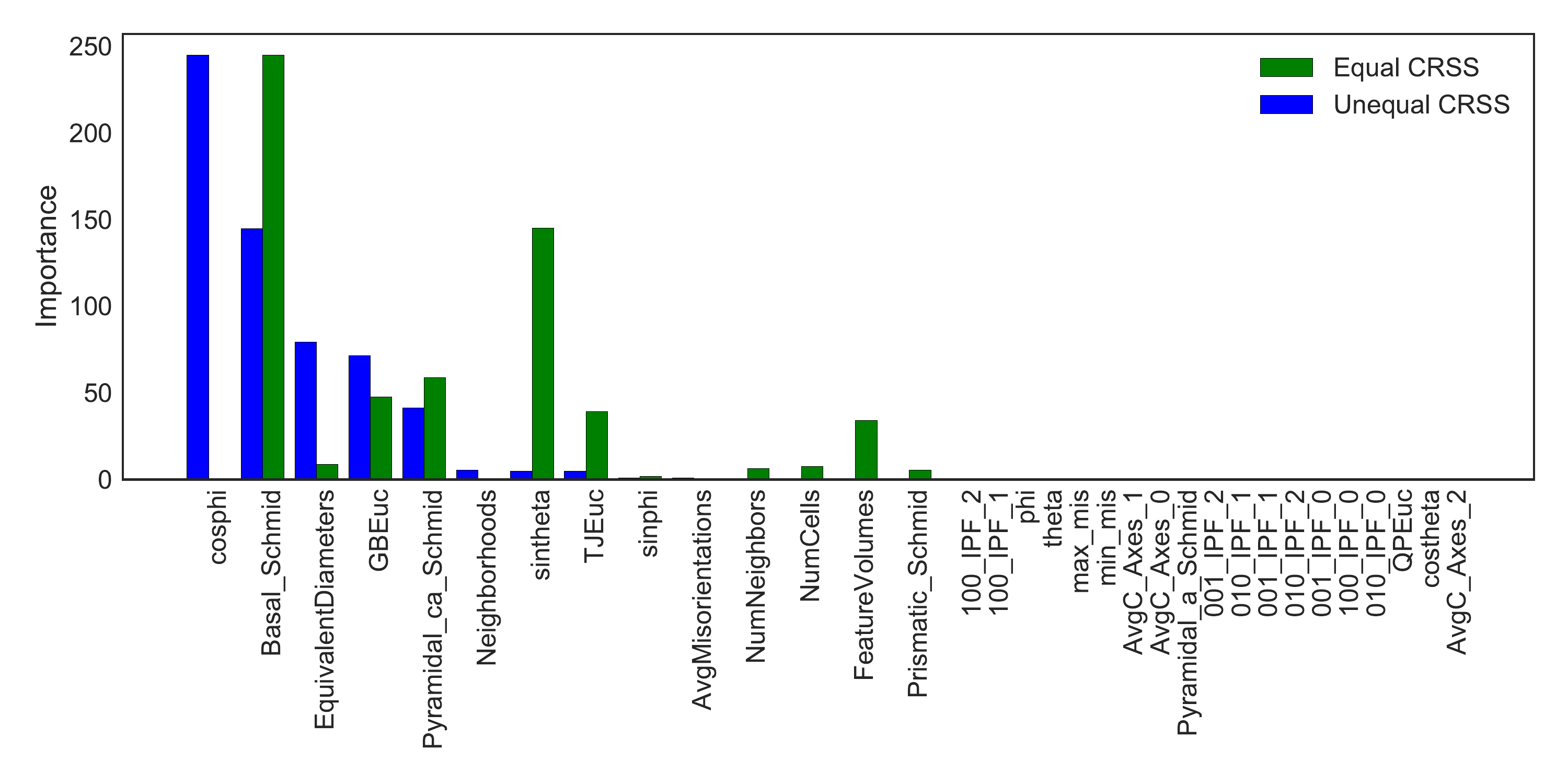}
    \caption{FeaLect variable importance in HCP materials showing selected features for Equal CRSS (green) and Unequal CRSS (blue) }
    \label{LassoHCP_new}
\end{figure}

Figure \ref{LassoHCP_new} shows the feature importances for the mixed-microstructure model calculated for the Equal CRSS and Unequal CRSS ratio cases using the FeaLect algorithm. In the case of an Equal CRSS ratio (green bars), the basal Schmid factor is the most important feature, followed by the HCP-c axis orientation ($sin\theta$) and the pyramidal $<a>$ Schmid factor. We calculated the Pearson correlation between the important features and stress hotspots (table \ref{PearsonHCP}), and found that hotspots tend to form in grains with higher polar and azimuthal angle of the HCP-c axis, which translates to grains with higher basal Schmid factor, which is proportional to the cosine of theta.  The elastic modulus for HCP materials (Ti) has a angular behavior which is captured by $\theta$. The elastic modulus is highest along $< 0001 >$ direction and lowest in the $[0001]$ plane. Hence hotspots form in grains with lower elastic modulus. This shows the power of feature selection to capture physical effects, since in the absence of heterogenous slip systems, the stress distribution is impacted by the directionality in elastic modulus which in turn is dependent on theta. This trend is similar to our result in FCC materials from \cite{Mangal2017b}; when the material has homogenous deformation modes, the most important features are those which couple the the loading direction and the crystallography. The geometry derived features come next on the feature importance plot, and we found that hotspots lie closer to grain boundaries, triple junctions and quadruple points i.e. form in smaller grains (table \ref{PearsonHCP}). This result is in agreement with \cite{Mangal2017b} and \cite{Rollett2010a} where stress hot spots were found to lie closer to microstructural features. 

For the Unequal CRSS ratio material, from figure \ref{LassoHCP_new} (blue bars), we see that the set of important features are HCP c-axis orientation (phi, theta, basal Schmid factor), grain size, pyramidal $<c+a>$ Schmid factor and shape averaged triple junction distance per grain. The top 3 important features include the grain size (equivalent diameter) in contrast to materials with homogenous deformation modes. From the Pearson correlation coefficients (table \ref{PearsonHCP}), we observe that hotspots lie closer to grain boundaries, triple junctions and quadruple points, form in grains with low basal and pyramidal $<c+a>$ Schmid factor and prefer a high prismatic $<a>$ Schmid factor. In the following section, we explore the effect of competing slip systems to better understand the feature importance results.

\begin{table}[!tb]
\begin{center}
\caption{Pearson Correlation Coefficients between features and Stress hotspots for HCP materials}
\label{PearsonHCP}
\begin{tabular}{*{5}{c}}
\toprule  \multirow{2}{*}{\textbf{Feature}} & \multicolumn{2}{c}{\textbf{Equal CRSS}} &  \multicolumn{2}{c}{\textbf{Unequal CRSS}} \\ 
& Correlation Coefficient & p-value & Correlation Coefficient & p-value  \\
\midrule
theta&0.002&0.27 &-0.0029&0.1083\\
phi&0.089&0.0 &0.1276&0.0\\
\midrule
Basal $<a>$ Schmid&0.5428&0.0 &-0.3933&0.0\\
Prismatic $<a>$Schmid&-0.5567&0.0 &0.490&0.0\\
Pyramidal $<a>$ Schmid&-0.0629&0.0 &0.490&0.0\\
Pyramidal $<c+a>$ Schmid&0.1181&0 &-0.1777&0.0\\
\midrule
GBEuc&-0.0027&0.14 &-0.0084&0.0\\
TJEuc&-0.0024&0.20 &-0.0094&0.0\\
QPEuc&-0.0021&0.25 &-0.0052&0.005\\
Equivalent Diameter&-0.0023 & 0.22 &-0.0087 & 0.0\\
\bottomrule
\end{tabular}
\end{center}
\end{table}

\subsection{Role of competing slip systems in stress hotspot formation}
\label{Role of competing slip systems in stress hotspot formation}
\begin{figure}[!tb]
    \centering
    \includegraphics[width=\textwidth]{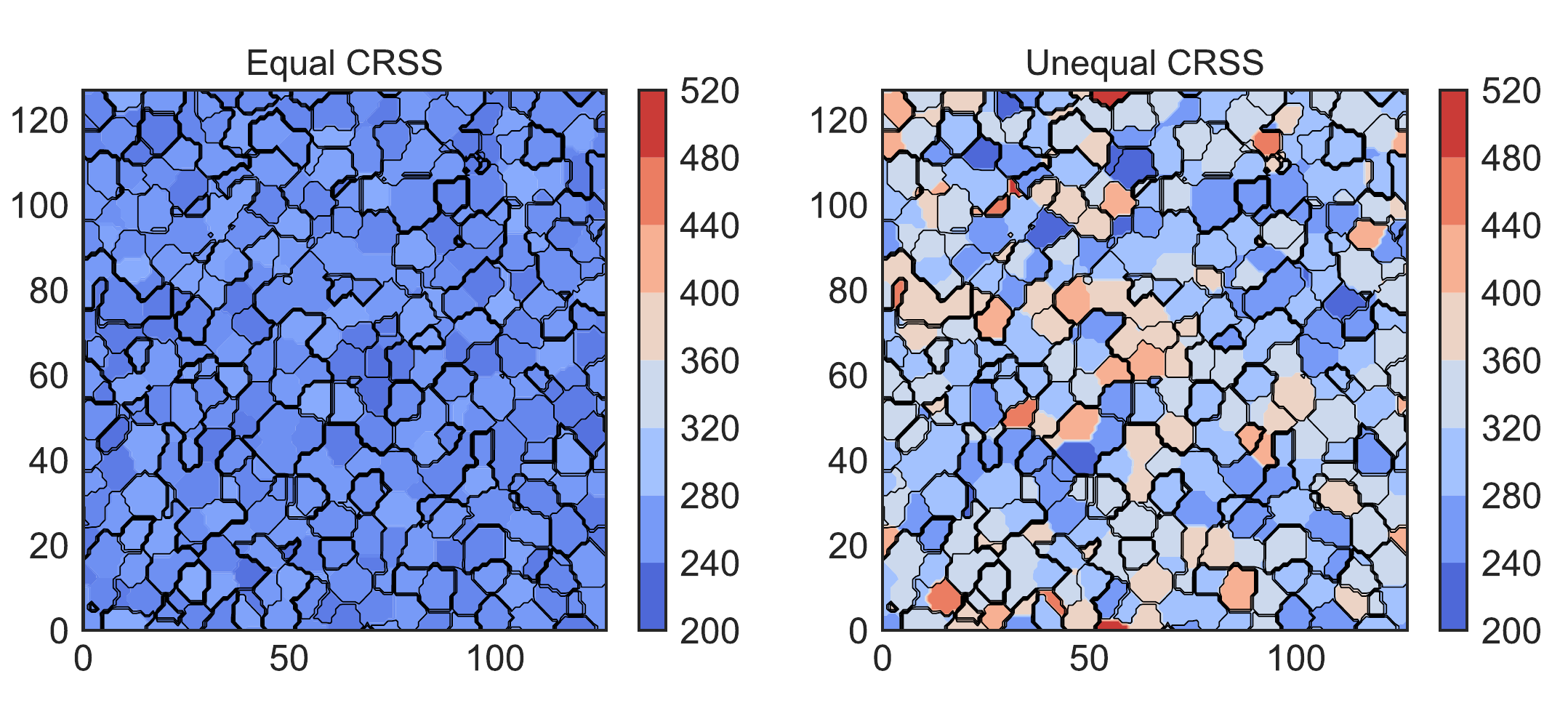}
    \caption{Cross section of a randomly textured 3-D equiaxed microstructure showing the Von Mises stress distribution under different SCYS topology regimes for a microstructure with random texture.}
    \label{fig:CRSS_ratio_crosssec}
\end{figure}

To compare the effect of the competing slip systems on stress hotspot formation in HCP materials, the set of microstructures with random texture is selected. Figure \ref{fig:CRSS_ratio_crosssec} shows the cross section of one of these microstructures, with the spatially resolved Von Mises stress field in the Equal and Unequal CRSS ratio cases. It can be observed that stress hotspots are more pronounced when a limited number of slip systems is available (Unequal CRSS), and for the same microstructure, hotspot location changes with available slip systems. It was found that the skewness of the grain averaged stress histogram for the Equal CRSS case is 0.085 and for the Unequal CRSS case it is an order of magnitude larger, 0.85; that is, when slip systems are limited, a heavy tailed stress distribution is observed. Due to the high CRSS for pyramidal $<c+a>$ slip compared to prismatic $<a>$ slip, some grains, due to their orientation, are at a disadvantage, because they cannot provide the necessary deformation modes required to close the yield surface. In such grains, the stress climbs very high and there is no clear yield, thus causing the heavy tail.

\begin{figure}[!tb]
    \centering
    \begin{subfigure}[t]{0.49\textwidth}
    \includegraphics[width=\textwidth]{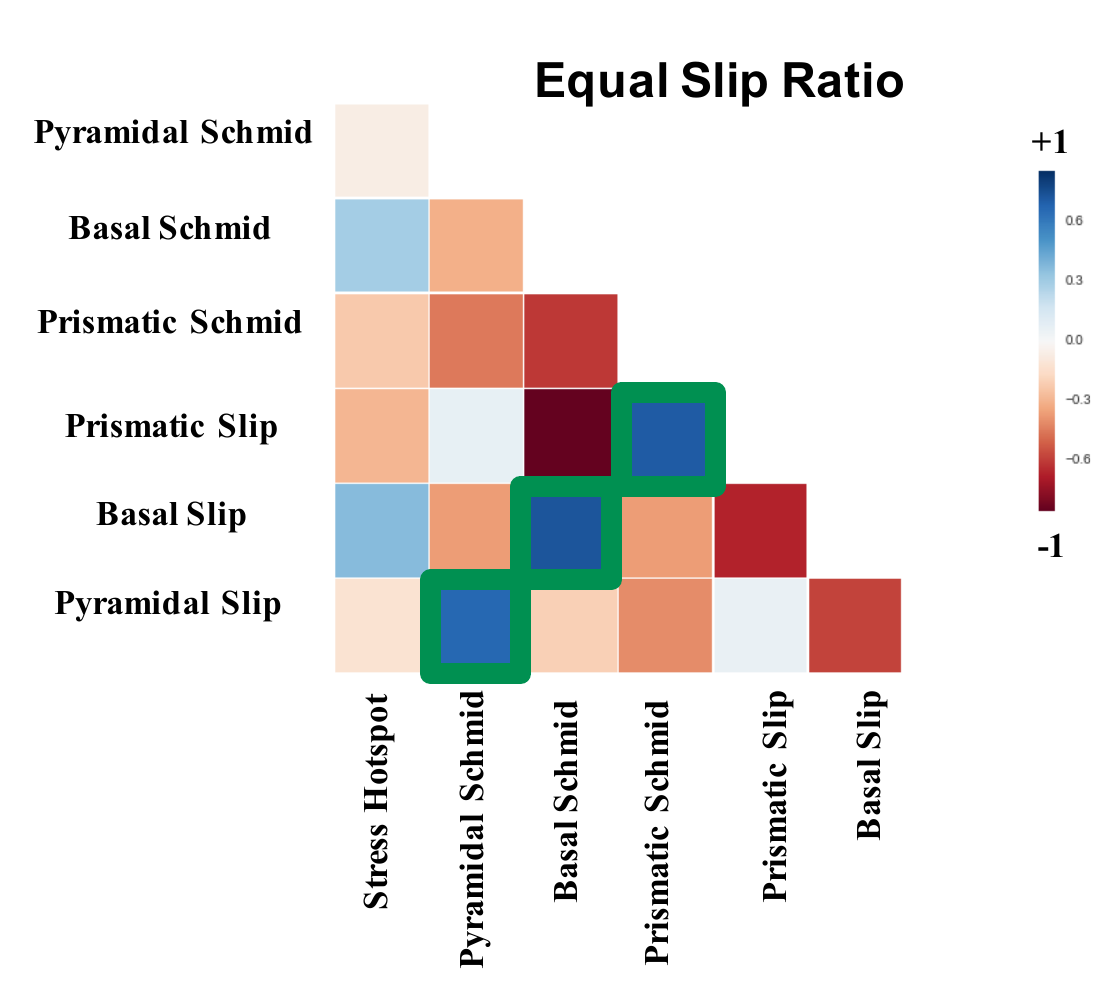}
    \caption{}
    \label{fig:CRSS_equal}
    \end{subfigure}
    \hfill
    \begin{subfigure}[t]{0.49\textwidth}
    \includegraphics[width=\textwidth]{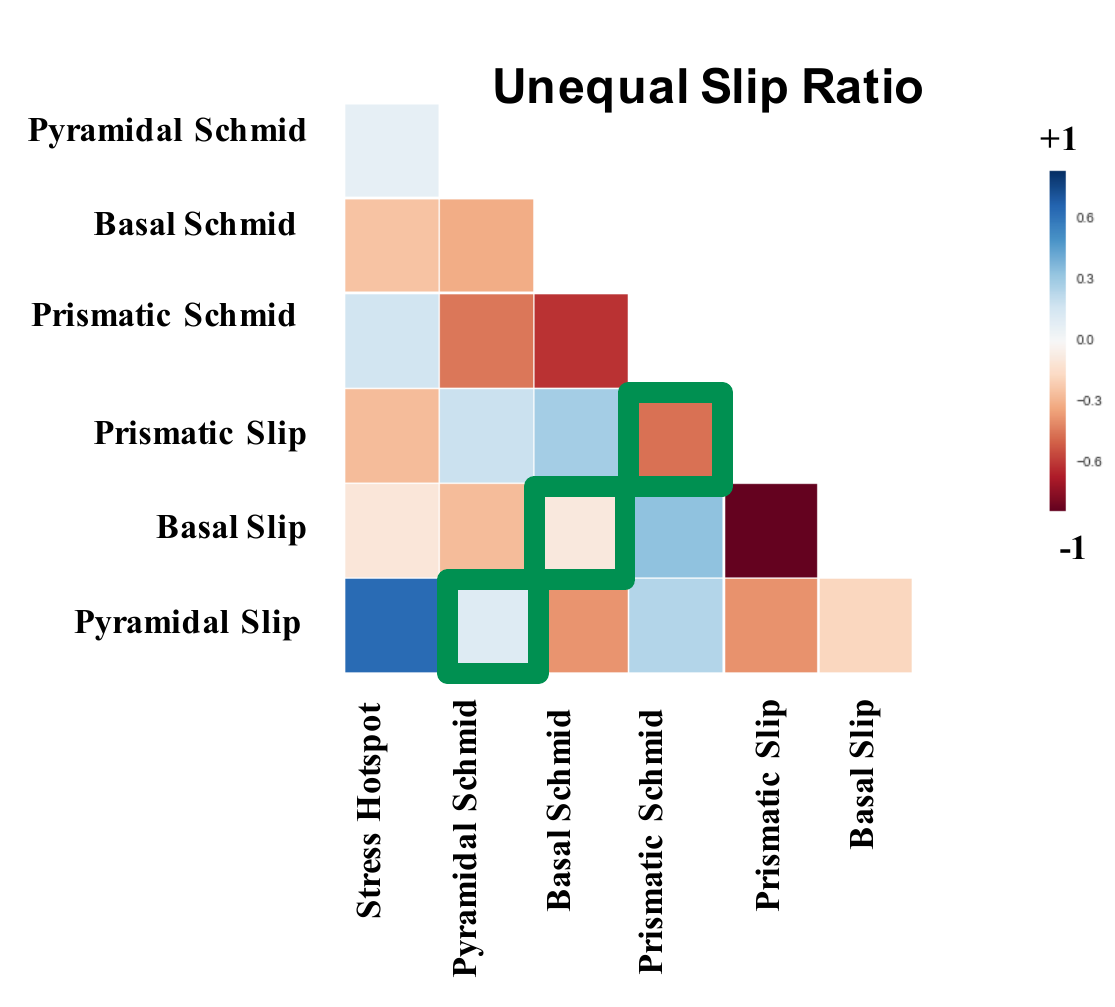}
    \caption{}
    \label{fig:CRSS_unequal}
    \end{subfigure}
    \caption{Correlation matrix for slip activities and Schmid factors (a) Equal CRSS ratio case and (b) Unequal CRSS ratio case. The correlation between corresponding slip activities and Schmid factors is highlighted. There is a strong positive correlation between pyramidal slip fraction and stress hotspot formation for the Unequal CRSS ratio case.}
\end{figure}

The grain population was sorted by stress values and divided into 10 bins, each having 10\% of the grains. Hence the last bin corresponds to the grains which are stress hotspots. The relative slip activities in each slip system were then compared between these bins, and the mean value of the slip activities in each bin was compared for the Equal and Unequal CRSS cases. We found that the Equal CRSS ratio hotspots have high basal slip fractions, whereas the Unequal CRSS ratio hotspots have higher pyramidal slip fraction. The number of active slip systems was found to be similar in both cases, thus following the single crystal yield surface (SCYS) criterion.

\begin{figure*}[htb]
\centering
    \begin{subfigure}[]{0.6\textwidth}
        \centering
        \includegraphics[width=\textwidth]{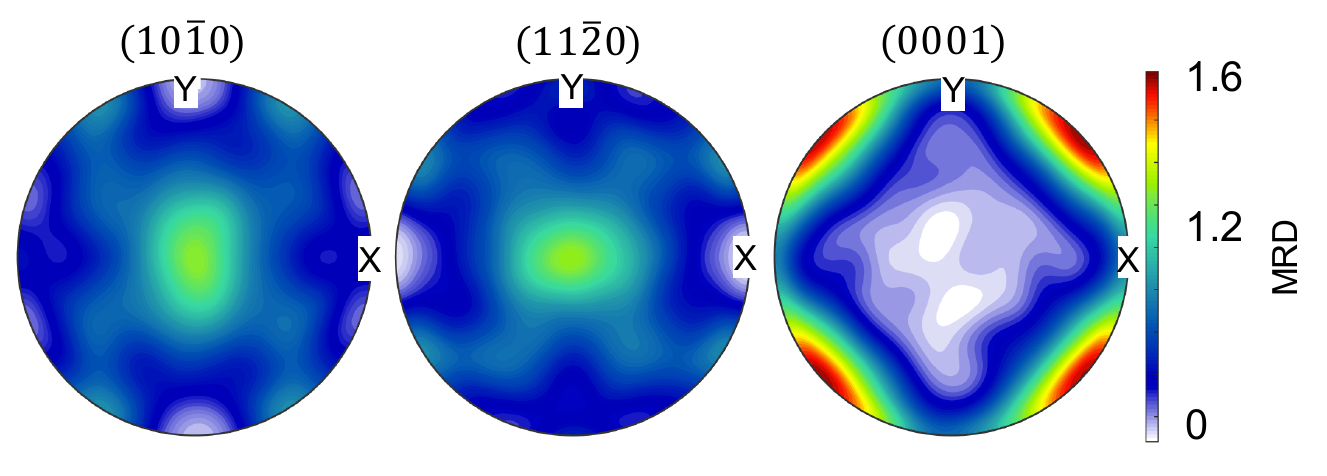}
        \caption{}
        \label{fig:CRSS_PF0}
    \end{subfigure}
    \\
    \begin{subfigure}[]{0.7\textwidth}
        \centering
        \includegraphics[width=\textwidth]{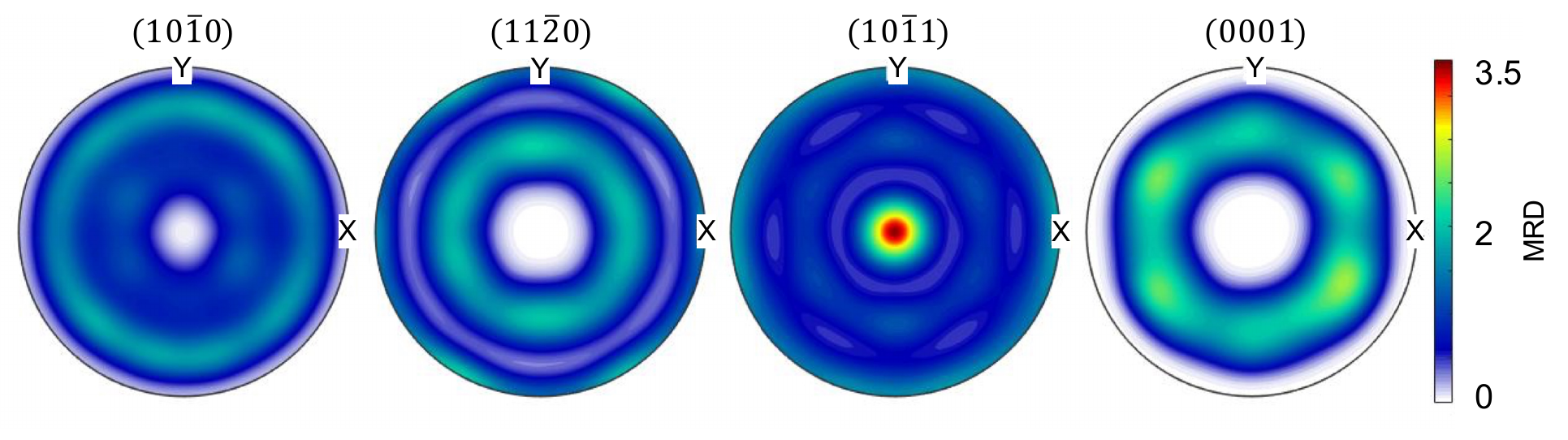}
        \caption{}
        \label{fig:CRSS_PFequal}
    \end{subfigure}
    \\
    \begin{subfigure}[]{0.7\textwidth}
        \centering
        \includegraphics[width=\textwidth]{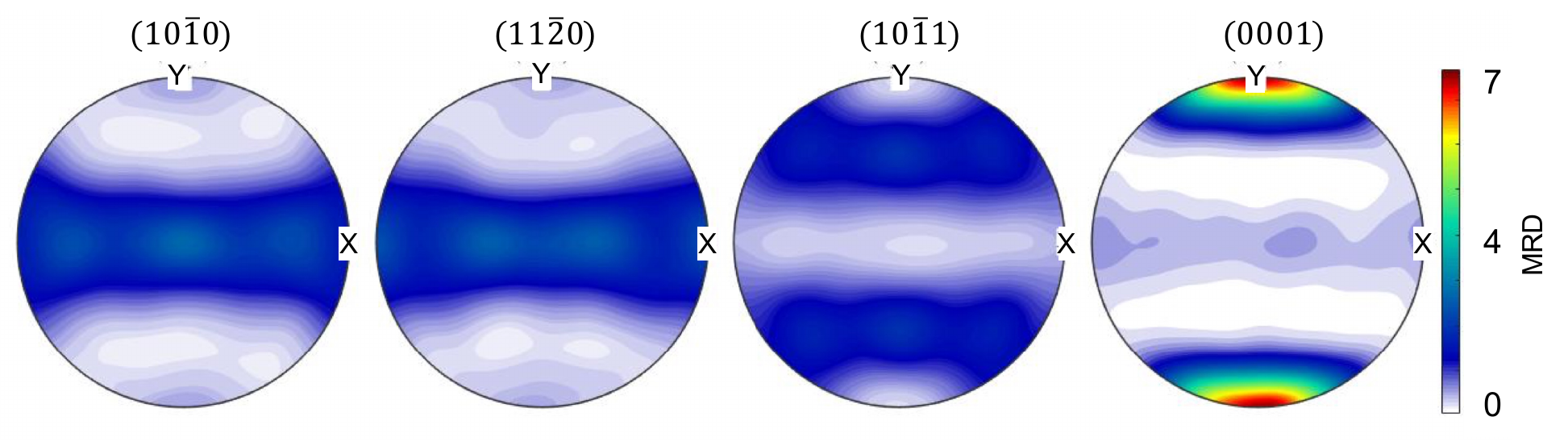}
        \caption{}
        \label{fig:CRSS_PFunequal}
    \end{subfigure}
     \caption{Pole figures showing texture of (a) starting random microstructure, (b) hot grains in the Equal CRSS case and (c) hot grains in the Unequal CRSS case. Note the different scale bars.}
    \label{fig:CRSS_PF}
\end{figure*}

To further understand the slip activities, a correlation matrix between the slip activity and the corresponding Schmid factors is plotted in figure \ref{fig:CRSS_equal} and \ref{fig:CRSS_unequal}. A strong positive correlation between slip activity and the corresponding Schmid factors is observed for the Equal CRSS ratio case, and the correlation is weak for the Unequal CRSS ratio case. The strong correlation in the equal slip case could be due to a more isotropic yield surface. Because the CRSS for all slip systems is the same, many slip systems are activated at the same stress, and the Schmid factor becomes important. In the Unequal CRSS case, the number of available slip systems is smaller, and even if the CRSS of a mode is very high, it might be activated to complete the yield surface to achieve 5 independent slip modes.

Figure \ref{fig:CRSS_PF} shows the pole figures for the starting microstructure and the hot grains. Starting with a random texture (figure \ref{fig:CRSS_PF0}), we notice that hotspots form in completely different textures in the two cases. 

For the Equal CRSS case, (figure \ref{fig:CRSS_PFequal}), it can be observed that the hotspot loading direction (the z-axis in the sample reference frame) has no preference to align with the $(10\bar{1}0)$ or $(11\bar{2}0)$ planes. However, the loading direction aligns with the [$10\bar{1}1$] pyramidal pole as seen from the (0001) and ($10\bar{1}1$) pole figure. When this happens, the loading direction lies in the pyramidal plane as shown in figure \ref{fig:EqualCRSSslips}. In this orientation, the Schmid factor favors prismatic slip, so if basal slip is becoming active in these grains, it should mean they have a higher stress.

For the Unequal CRSS case, from the set of hotspot pole figures (figure \ref{fig:CRSS_PFunequal}), we can see that there is no preference for the loading direction to align with the prismatic and pyramidal planes. The c-axis aligns with the sample y-axis which means these grains have a low elastic modulus. Since the c-axis is perpendicular to the tensile axis, the deformation along the tensile direction can be accommodated by prismatic slip, and if pyramidal slip is occurring, it requires a very high stress. From the comparison of pole figures of hot grains, the dominant slip modes cannot be predicted with confidence.  

\section{Conclusions}
\begin{itemize}[leftmargin=*]
    \item Stress hotspots can be predicted with 82.5\% AUC in HCP materials with Equal CRSS ratio, and 81.18\% AUC in HCP materials with Unequal CRSS ratio using random forest models. We observe that the performance of  Mixed-models is comparable to or better than Partition-models. This could mean the existence of common factors independent of the macro-texture which cause stress hotspots in a material. 
        \item A change in material composition will result in altered constitutive parameters, and consequently, the mechanical response. This changes the microstructural descriptors needed, and hence models need to be built for each material.  
    \item Contrasting stress hotspot formation for Equal vs Unequal CRSS ratios in materials with random texture, we observe: 
    \begin{itemize}[leftmargin=*]
    	\item Stress hotspots are more pronounced when a limited number of slip systems is available (Unequal CRSS), and for the same microstructure, hotspot location changes with available slip systems. 
        \item Stress hotspots in the Equal CRSS ratio case have high basal slip fractions and strong positive correlation between slip activity and corresponding Schmid factors, which could be due to an isotropic yield surface. 
        \item Stress hotspots in the Unequal CRSS ratio case have higher pyramidal slip fraction and weak correlations between corresponding slip activities and Schmid factors. This could be due to the limited number of slip systems.
        \item A comparison between the feature importance results reveals that the macro-texture (HCP c-axis orientation) mainly determines stress hotspots in the Equal CRSS ratio case. In the Unequal CRSS ratio case, both crystallography and geometry based features are required to predict stress hotspots. 
    \end{itemize}
   \end{itemize}

\subsection{Contributions}
We have successfully demonstrated the applicability of a data driven approach for predicting stress hotspots in different kinds of HCP materials. Using feature importance plots, we are able to gain objective insights on how hotspot formation varies with material parameters such as CRSS ratio. Hence the framework used in this work is not limited to predicting stress hotspots in HCP materials, but can be extended to various polycrystalline materials, and a wide range of structure-property relationships in materials.

\section{Acknowledgements }
This work was performed at Carnegie Mellon University and has been supported by the United States National Science Foundation award number DMR-1307138 and DMR-1507830.

\section*{References}

\bibliography{library}

\setcounter{equation}{0}
\setcounter{figure}{0}
\setcounter{table}{0}
\setcounter{page}{1}
\setcounter{section}{0}
\makeatletter
\renewcommand{\theequation}{S\arabic{equation}}
\renewcommand{\thefigure}{S\arabic{figure}}
\renewcommand{\thetable}{S\arabic{table}}
\renewcommand{\bibnumfmt}[1]{[S#1]}
\renewcommand{\citenumfont}[1]{S#1}

\appendix
\section{Constitutive Parameters: Hexagonal Close Packed Materials}
\label{AppendixA} 
The constitutive model parameters for HCP materials are similar to a general alpha-titanium alloy having an equiaxed microstructure \cite{Ikehata2004}. The single crystal elastic constants are given in table \ref{HCPelastic1}. Only three slip systems are considered: basal $\{0001\}[11\overline{2}0]$, prismatic $\{10\overline{1}0\}[11\overline{2}]$ and pyramidal $<c+a>$. Two cases are considered based on the strength of different slip systems i.e.  having Equal and Unequal CRSS ratios. The Equal CRSS case is hypothetical and is analyzed purely for model development and analysis. The second case with the CRSS ratio of basal$<a>$: prismatic$<a>$: pyramidal$<c+a>$ = 1: 0.7 : 3 has the same single crystal elastic stiffness constants (table \ref{HCPelastic1}). The boundary conditions correspond to uniaxial tension along Z, with an applied strain rate component along the tensile axis $\dot{\epsilon_{33}} = 1s^{-1}$. The EVPFFT simulation was carried out in 200 steps of 0.01\%, up to a strain of 2\%. \\

\begin{figure}[!htbp]
\centering        
            \includegraphics[width=0.5\textwidth]{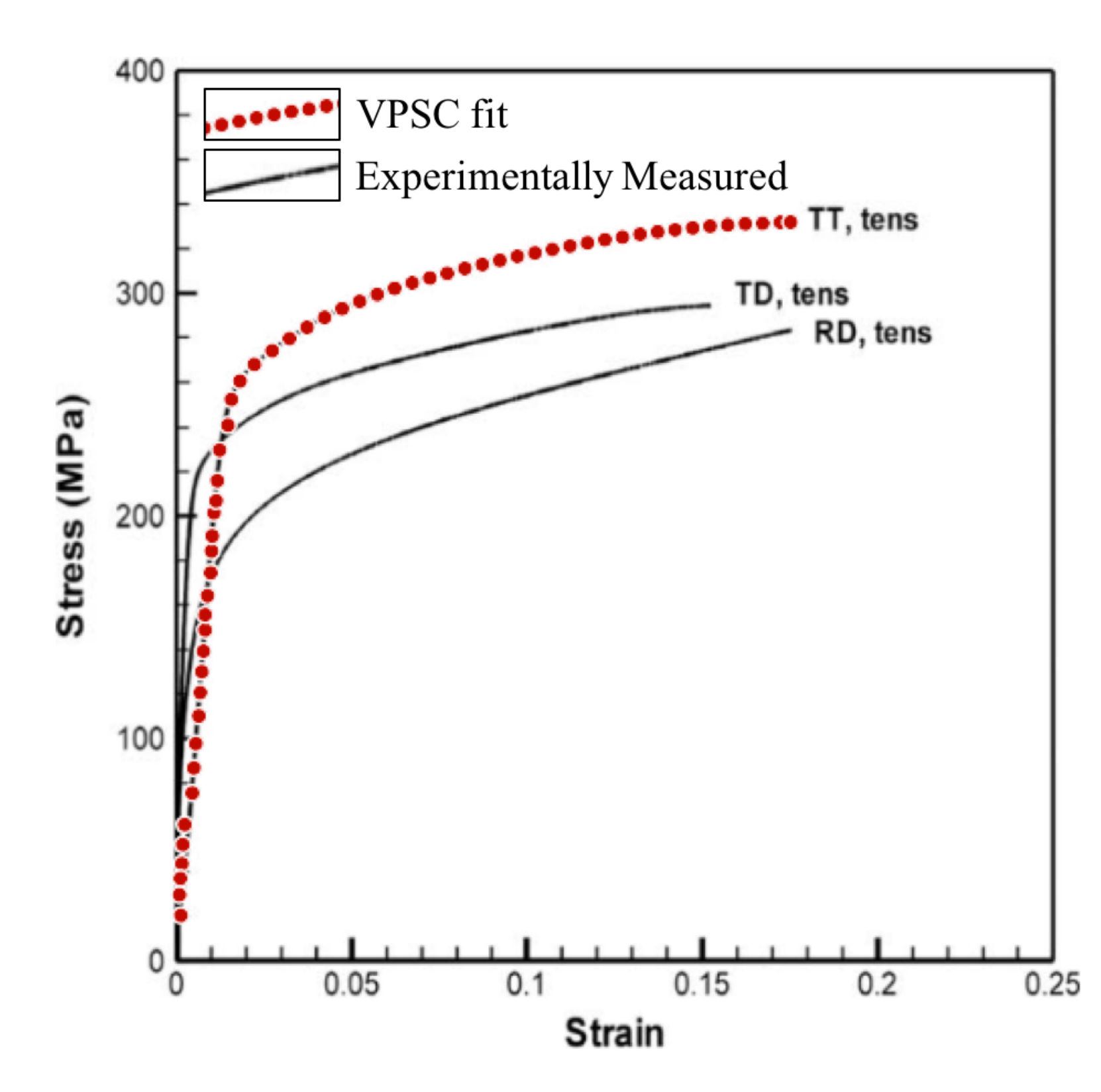}
            \caption{HCP $\alpha$-Titanium fit}
            \label{fig:AlphaTi}
\caption{VPSC simulation fit to the experimentally observed stress-strain curve for alpha-Titanium.}
\label{fig:VPSCfit}
\end{figure}

To obtain the actual CRSS values and the Voce hardening parameters, the Voce model was fit to an experimentally measured stress- strain curve for uniaxial tension in $\alpha$-Titanium \cite{Nixon2010} using the VPSC formulation. The results of the fitting are shown in figure \ref{fig:AlphaTi}, and table \ref{Voce} lists the CRSS values and hardening parameters obtained for each CRSS ratio. Note that, for HCP materials; we have used 8 different kinds of textures summarized in figure \ref{fig:HCPtex}. The stress exponent is 10 for all cases. \\

\begin{figure*}[!htbp]
 \centering
 \includegraphics[width=0.8\textwidth]{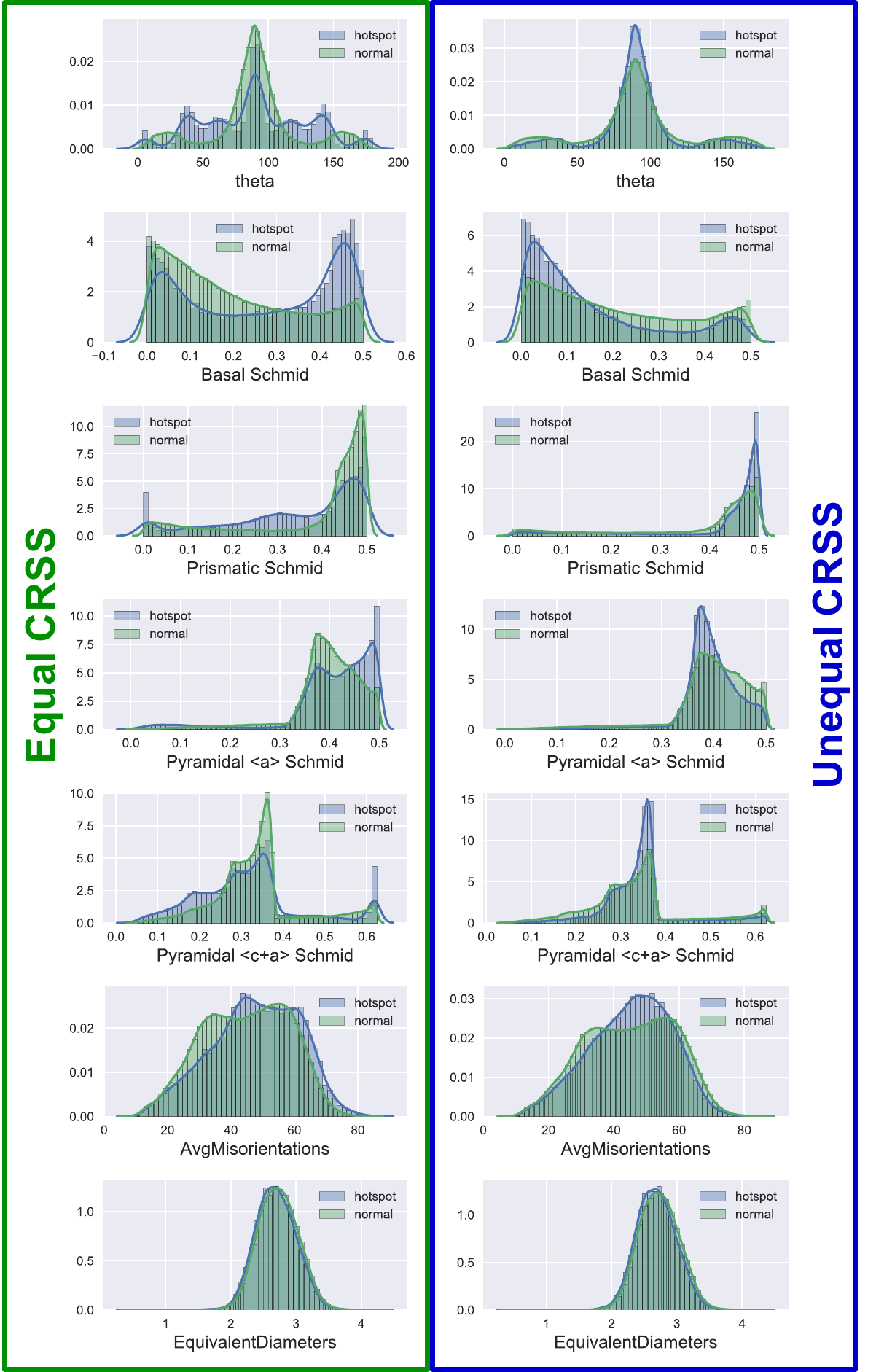}
 \caption{Histograms of some important features distinguishing hot and normal grains in the Equal CRSS and Unequal CRSS materials.}
 \label{fig:HCPfeaturedist}
\end{figure*}

To understand how the most predictive features influence hotspot formation in HCP materials, the distribution of these feature values in normal and hot grains is plotted as shown in figure \ref{fig:HCPfeaturedist} for both kinds of materials. Feature distributions for the Equal CRSS materials are in the first column and the Unequal CRSS materials are in the second column.

\textbf{Equal CRSS: }
From the plot for theta for hot grains, we can see that there is a peak at high theta values, where the elastic modulus is low i.e. undergoing more plastic deformation. There are three smaller peaks near 64, 40 and 8 degrees, which might be due to more complex effects of plasticity.

\textbf{Unequal CRSS: }
The distribution of different Schmid factors between hot and normal grains for this material is very different from the Equal CRSS material. But in both cases, stress hotspots tend to form in grains with higher average misorientation and grains with lower elastic modulus. Although there is no visible difference in grain sizes between hot and normal grains, this feature becomes distinguishing in association with texture derived features.

\begin{table}[!bt]
\centering
\caption{Voce Hardening law parameters for $\alpha$-Titanium}%
\label{Voce}
\small
\begin{tabular}{p{1.1cm} p{1.6cm} p{1cm} p{1cm}p{0.8cm}p{0.8cm}}
\toprule 
\textbf{CRSS ratio}& \textbf{Slip System} & $\mathbf{\tau_0^s}$ (MPa) & $\mathbf{\tau_1^s}$ (MPa) & $\mathbf{\theta_0^s}$ & $\mathbf{\theta_1^s}$\\ 
\midrule
\multirow{3}{*}{0.7:1:3 }&Basal& 82.8 & 36.7  & \multirow{3}{*}{406.3} & \multirow{3}{*}{4.6}\\
&Prismatic& 57.9  & 25.7  \\
&Pyramidal& 248.5  & 110.1  \\
\midrule
 1:1:1 & All & 100& 50 & 500 & 10 \\
\bottomrule
\end{tabular}
\end{table}

\begin{table}[!bt]
\centering
\caption{Single crystal elastic stiffness constants  (in GPa)}
\label{HCPelastic1}
\small
\begin{tabular}{p{2cm} p{0.5cm} p{0.5cm} p{0.5cm} p{0.5cm} p{0.5cm} p{0.5cm}}
\toprule 
\multicolumn{1}{c}{Material} & \multicolumn{1}{c}{$\mathbf{C_{11}}$} & \multicolumn{1}{c}{$\mathbf{C_{12}}$} & \multicolumn{1}{c}{$\mathbf{C_{13}}$} & \multicolumn{1}{c}{$\mathbf{C_{33}}$} & \multicolumn{1}{c}{$\mathbf{C_{44}}$} & \multicolumn{1}{c}{$\mathbf{C_{66}}$}\\ \midrule
$\alpha$-Titanium (approx.)& 170 & 98 & 86 & 204 & 51 & 66 \\ 
\bottomrule 
\end{tabular}
\end{table}

%
%

\clearpage
\section{Geometric and crystallographic descriptors used for machine learning} \label{Featurenames}
\begin{table}[!htb]
\centering
\caption{Feature name descriptions}
\small
\begin{tabular}{|p{0.18\linewidth}p{0.3\linewidth}|p{0.18\linewidth}p{0.3\linewidth}|}
\toprule 
\textbf{Feature name Abbreviation} & \textbf{Description} & \textbf{Feature name Abbreviation} & \textbf{Description}\\ \midrule
Schmid\_0 & Basal $<a>$ Schmid factor & 100\_IPF\_x & Distance from the corners of the 100 Inverse pole figure\\
\midrule
Schmid\_1 & Prismatic $<a>$ Schmid factor & 001\_IPF\_x & Distance from the corners of the 001 Inverse pole figure\\
\midrule
Schmid\_2 & Pyramidal $<a>$ Schmid factor & AvgC\_Axes\_x & Unit vector components describing the c axis orientation for hcp\\
\midrule
Schmid\_3 & Pyramidal $<c+a>$ Schmid factor & Max\_mis & Maximum misorientation between a grain and its nearest neighbor\\
\midrule
Surface area volume ratio & Ratio between surface area and volume of a grain & Min\_mis & Minimum misorientation between a grain and its nearest neighbor\\
\midrule
theta & Polar angle of hcp c axis w.r.t sample frame & AvgMisorientations & Average misorientation between a grain and its nearest neighbor\\
\midrule
phi & Azimuthal Angle of hcp c axis w.r.t. sample frame & QPEuc & Average distance of a grain to quadruple junctions\\
\midrule
TJEuc & Average distance of a grain to triple junctions & NumNeighbors & Number of nearest neighbors of a grain\\
\midrule
GBEuc & Average distance of a grain to grain boundaries & Neighborhoods & Number of grains having their centroid within the 1 multiple of equivalent sphere diameters from each grain\\
\midrule
KernelAvg & Average misorientation within a grain & FeatureVolumes & Volume of grain\\
\midrule
Omega3s & 3rd invariant of the second-order moment matrix for the grain, without assuming a shape type & Equivalent Diameters & Equivalent spherical diameter of a grain\\
\midrule
Surface Features & 1 if grain touches the periodic boundary else 0 & AspectRatios & Ratio of axis lengths (b\/a and c\/a) for best-fit ellipsoid to grain shape\\
\bottomrule
\end{tabular}
\end{table}

\end{document}